\documentclass{article}
\def\~{\tilde}
\usepackage{graphicx}

\begin{document}
\title{Reply to: Comment on ``Bohmian prediction about a two double-slit experiment and its disagreement with SQM"}
\author{O. Akhavan \footnote{E-mail: akhavan@mehr.sharif.edu} and M. Golshani
\footnote{E-mail: golshani@ihcs.ac.ir}}
\date{\small{\it Department of Physics, Sharif University of Technology,
P.O. Box 11365--9161, Tehran,  Iran\\Institute for Studies in
Theoretical Physics and Mathematics, P.O. Box 19395--5531, Tehran,
Iran}} \maketitle

\begin{abstract}
In a recent paper, Struyve {\it et al.} [Struyve W, De Baere W, De
Neve J and De Weirdt S 2003 {\it J. Phys. A} {\bf 36} 1525]
attempted to show that the thought experiment proposed in
[Golshani M and Akhavan O 2001 {\it J. Phys. A} {\bf 34} 5259]
cannot distinguish between standard and Bohmian quantum mechanics.
Here, we want to show that, in spite of their objection, our
conclusion still holds out.
\end{abstract}

PACS number(s): 03.65.Bz\\

Recently, a thought experiment was proposed in {\cite{Gol}} to
show that the standard quantum mechanics (SQM) can present
incompatible predictions as compared with Bohmian quantum
mechanics (BQM) at the individual level of particles.  But, this
proposal was diligently criticized by Struyve {\it et al.}
{\cite{Str}} by resorting to the quantum equilibrium hypothesis
(QEH).  It seems useful, before any discussion on the raised
objection, to outline the proposed experiment.

Consider an original type of EPR source {\cite{Ein}} which emits a
pair of identical non-relativistic particles labelled by 1 and 2
with vanishing total momentum.  The source is placed at the origin
of the $x-y$ coordinate system which is taken to be the
geometrical center of two identical double-slit screen in parallel
to the $y$-axis.  The slits are labelled by $A$ and $B$ as well as
$A'$ and $B'$ on the right and left screens, respectively.  They
have the coordinates $(\pm d,\pm Y)$ with the half-width of each
slit being $\sigma_0$.  For simplicity, let us suppose that there
is just one pair of particles in the system at any moment.  Thus,
the general wave function of the system describing the two
entangled particles emerging from the slits can be written as
\begin{eqnarray}
\psi(x_{1},y_{1};x_{2},y_{2};t)&=&N[\psi_{A}(x_{1},y_{1},t)\psi_{B^{'}}(x_{2},y_{2},t)\pm\psi_{A}(x_{2},y_{2},t)\psi_{B^{'}}(x_{1},y_{1},t)\nonumber\\
&&+\psi_{B}(x_{1},y_{1},t)\psi_{A^{'}}(x_{2},y_{2},t)\pm\psi_{B}(x_{2},y_{2},t)\psi_{A^{'}}(x_{1},y_{1},t)]
\end{eqnarray}
where $N$ is a renormalization constant and the upper (lower) sign
refers to bosonic (fermionic) property of the two-particle system.
Furthermore, it is assumed that the slits produce Gaussian wave
functions in the form
\begin{eqnarray}{\label{2}}
\psi_{A,B}(x,y,t)&=&(2\pi \sigma_{t}^{2})^{-1/4}exp[-(\pm
y-Y-\hbar
k_{y}t/m)^{2}/4\sigma_{0}\sigma_{t}]\nonumber\\
&&\times exp[i(k_{x}(x-d-\hbar k_{x}t/2m)+k_{y}(\pm y-Y-\hbar
k_{y}t/2m)]
\end{eqnarray}
where the index $A (B)$ is related to the upper (lower) sign, and
\begin{equation}
\sigma_{t}=\sigma_{0}(1+\frac{i\hbar t}{2m\sigma_{0}^{2}}).
\end{equation}
Notice that, the form of $\psi_{A',B'}$ is the same as
$\psi_{A,B}$ with the parameter $d$ being replaced by $-d$ in
Eq.~(\ref{2}).  A joint detection of the two particles is
simultaneously done on two screens placed at $\pm (D+d)$,
perpendicular to the $x$-axis.  It is well known that, based on
SQM, the probability of the joint detection of a pair, at
positions $y_1=Q_1$ and $y_2=Q_2$ on the two screens, is
\begin{equation}
P_{12}(Q_1,Q_2,t_0)=\int_{Q_1}^{Q_1+\triangle
Q}\int_{Q_2}^{Q_2+\triangle Q}|\psi(y_{1},y_{2},t_0)|^{2}dy_1dy_2
\end{equation}
where $t_0=mD/\hbar k_x$ and $\triangle Q$ is the size of position
detectors.  On the other hand, according to BQM, one can show that
{\cite{Gol}}
\begin{equation}
y(t)=y(0)\sqrt{1+(\hbar t/2m\sigma_{0}^{2})^{2}}
\end{equation}
where $y(t)=(y_1(t)+y_2(t))/2$ is center of mass position of the
two particles in the $y$-direction at time $t$.  Now, if we can
adjust $y(0)=0$, then BQM predicts that each pair of particles
must be observed symmetrically with respect to the $x$-axis on the
screens.  In addition, it predicts that detection of the two
particles at one side of the $x$-axis on the screens is
impossible.  But, according to SQM, the probability of
asymmetrical joint detection of the two particles or finding them
on one side of the $x$-axis can be non-zero, contrary to BQM's
prediction.  It can easily be seen that, the difference between
the two theories is rooted in the position entanglement condition
$y(0)=0$.  In fact, Struyve {\it et al.} {\cite{Str}} believe that
by considering QEH, this initial entangled condition is not
realizable in BQM.  Thus, in the following, the discussion is
mainly concentrated on this issue.

Let us begin with some more detailed description of the
experiment.  Suppose that before the arrival of the two particles
on the slits, the entangled wave function describing them is given
by
\begin{eqnarray}{\label{5}}
\psi_0(x_1,y_1;x_2,y_2)&=&\chi(x_1,x_2)\hbar\int_{-\infty}^{+\infty}exp[ik_y(y_1-y_2)]dk_y\nonumber\\
&=&2\pi\hbar\chi(x_1,x_2)\delta(y_1-y_2)
\end{eqnarray}
where $\chi(x_1,x_2)$ is the $x$-component of the wave function
that could have a form similar to the $y$-component.  However, its
form is not important for the present work.  The wave function
~(\ref{5}) is just the one represented in {\cite{Ein}}, and it
shows that the two particles have vanishing total momentum in the
$y$-direction, and their $y$-component of the center of mass is
exactly located on the $x$-axis.  This is not inconsistent with
Heisenberg's uncertainty principle, because
\begin{eqnarray}{\label{H}}
[p_{y_1}+p_{y_2},y_1-y_2]=0.
\end{eqnarray}
Furthermore, the considered source is not necessarily a point
source, and the two entangled particles are uniformly distributed
in the $y$-direction.

We have seen that, the wave function ~(\ref{5}) implies that there
are initial position and momentum entanglements for the pair in
the $y$-direction. Now, it is interesting to know what can happen
to these entanglements when the two particles emerge from the
slits. We assumed that the slits produce the Gaussian wave
functions represented by Eq.~(\ref{2}). Thus, when the particles
pass through the slits, the transformation
\begin{eqnarray}{\label{6}}
\psi_0(x_1,y_1;x_2,y_2)\longrightarrow \psi(x_1,y_1;x_2,y_2)
\end{eqnarray}
occurs to the wave function describing the system.  Now, there is
a question as to whether the position entanglement property of the
two particles is kept after this transformation.  To answer this
question, one can first examine the effect of the total momentum
operator on the wave function of the system,
$\psi(x_1,y_1;x_2,y_2;t)$, which yields
\begin{eqnarray}{\label{7}}
(p_{y_1}+p_{y_2})\psi(x_1,y_1;x_2,y_2;t)&=&-i\hbar
(\frac{\partial}{\partial y_1}+\frac{\partial}{\partial
y_2})\psi(x_1,y_1;x_2,y_2;t)\nonumber\\
&=&i\hbar
(\frac{y_1(t)+y_2(t)}{2\sigma_0\sigma_t})\psi(x_1,y_1;x_2,y_2;t)
\end{eqnarray}
where we see that the wave function is an eigenfunction of the
total momentum operator.  Now, if we can assume that the total
momentum of the particles is remained zero at all times (an
assumption about which we shall elaborate later on), then it can
be concluded that the center of mass of the two particles in the
$y$-direction is always located on the $x$-axis.  In other words,
a momentum entanglement in the form $p_1+p_2=0$ leads to the
position entanglement in this experiment. However, Born's
probability principle, i.e. $P=|\psi|^2$, which is a basic rule in
SQM, shows that the probability of asymmetrical joint detection of
the two particles can be non-zero on the screens. Thus, there is
no position entanglement and consequently no momentum entanglement
between the two particles.  This compels us to believe that,
according to SQM, the momentum entanglement of the two particles
must be erased during their passage through the slits, and the
center of mass position has to be distributed according to
$|\psi|^2$.

In BQM, however, Born's probability principle is not so important
as a primary rule and all particles follow well-defined tracks
determined by the wave function  $\psi({\bf{x}},t)$, using the
guidance condition
\begin{eqnarray}{\label{8}}
{\bf{\dot{x}}}_i({\bf{x}},t)=\frac{\hbar}{m_i}Im(\frac{\nabla_{i}\psi({\bf{x}},t)}{\psi({\bf{x}},t)})
\end{eqnarray}
with the unitary time development governed by
Schr$\ddot{o}$dinger's equation.  Now, let us review the previous
details, but this time in BQM frame.  Based on our supposed EPR
source, there are momentum and position entanglements between the
two particles before they were arrived on the slits.  Then, the
wave function of the emerging particles from the slits suffers a
transformation represented by Eq. ~({\ref{6}}).  It is not
necessary to know in details how this transformation acts, but
what is important is that the two double-slit screens are
considered to be completely identical. Thus, we expect that the
two particles in the slits undergo the same transformation(s), and
so the momentum entanglement, i.e. $p_1+p_2=0$, must be still
valid in BQM which is a deterministic theory, contrary to SQM.
Then, according to Eq. ~({\ref{7}}), the validity of the momentum
entanglement immediately leads to the position entanglement
\begin{eqnarray}{\label{9}}
y(t)=\frac{1}{2}(y_1(t)+y_2(t))=0.
\end{eqnarray}
We would like to point out that this entanglement is obtained by
using the quantum wave function of the system.  Therefore, the
claim that the supposed position entanglement can not be
understood by using the assumed wave function for the system is
not correct.  By the way, if we accepted that the momentum
entanglement is not kept and consequently $y(0)$ obeys QEH, then
deterministic property of BQM, which is a main property of this
theory, must be withdrawn.  However, it is well known that Bohm
{\cite{Bohm}} put QEH only as a subsidiary constraint to ensure
the consistency of the motion of an ensemble of particles with
SQM's results.  Thus, although in this experiment, the center of
mass position of the two entangled particles turns out to be a
constant in BQM frame, the position of each particle is consistent
with QEH so that the final interference pattern is identical to
what is predicted by SQM. Therefore, Bohm's aim concerning QEH is
still satisfied and deterministic property of BQM is left intact.
In addition, superluminal signals resulting from nonlocal
conditions between our distant entangled particles are precisely
masked by considering QEH for the distribution of each entangled
particle.

So far, contrary to the belief of Struyve {\it et al.}
{\cite{Str}} concerning QEH, we have shown that in BQM frame, the
center of mass position of the two entangled particles can be
considered to be a constant, without any distribution.  So, this
novel property provides a way to make a discrepancy between SQM
and BQM, even for an ensemble of entangled particles.  For
instance, suppose that we only consider those pairs one of which
arrives at the upper half of the right screen.  Thus, BQM predicts
that only detectors located on the lower half of the left screen
become ON and the other ones are always OFF.  In fact, we obtain
two identical interference pattern at the upper half of the right
screen and the lower half of the left screen.  Instead, SQM is
either silent or predicts a diluted interference pattern at the
left screen. Therefore, concerning to the validity of the initial
constraint $y(0)=0$ in BQM, selecting of some pairs to obtain a
desired pattern, which is called selective detection in
{\cite{Gol}}, can be applied to evaluate the two theories, at the
ensemble level of pairs.

In conclusion, the reason for the existence of these differences
between SQM and BQM in this thought experiment is that, in BQM as
a deterministic theory, the position and momentum entanglements
are kept at the slits, while in SQM, due to its probabilistic
interpretation, we must inevitably accept that the entanglements
of the two particles are erased when the two particles pass
through the slits. Meantime, the saved position entanglement in
BQM, i.e. $y(0)=0$, which is a result of the deterministic
property of the theory is not inconsistent with QEH, because we
are still able to reproduce SQM's prediction for an ensemble of
such particles, just as QEH requires. Therefore, our proposed
experiment is still a suitable candidate to distinguish between
the standard and Bohmian quantum mechanics. Furthermore, it is
worthy to note that, based on a recent work on Bohmian
trajectories for photons {\cite{Ghose}}, the first effort for the
realization of this type of experiment was performed by Brida {\it
et al.} {\cite{Brida}} very recently, using correlated photons
produced in type I parametric down conversion. This can stimulate
more serious and interesting discussions on the possible
incompatibilities between SQM and BQM, both theoretically and
experimentally.

\end{document}